\documentclass[9pt,twocolumn,twoside]{pnas-new}

\templatetype{pnasresearcharticle} 
\usepackage{blindtext}
\usepackage{graphicx}
\usepackage[utf8]{inputenc}

\usepackage{epstopdf, epsfig}
\usepackage{amsmath}
\usepackage{float}
\usepackage{stmaryrd}
\usepackage{braket}
\usepackage{xcolor,ulem}

\usepackage{bm}
\newcommand {\tcr}{\textcolor{black}}

\setboolean{displaywatermark}{false}

\begin{document}

\title{Thermoelectricity  at a gallium-mercury liquid metal interface}

\author[1]{Marlone Vernet}
\author[2]{Stephan Fauve}
\author[3,4]{Christophe Gissinger}

\affil[1]{Laboratoire de Physique de l'ENS, ENS, UPMC, CNRS; 24 rue Lhomond, 75005 Paris, France}
\affil[2]{Laboratoire de Physique de l'ENS, ENS, UPMC, CNRS; 24 rue Lhomond, 75005 Paris, France}
\affil[3]{Laboratoire de Physique de l'ENS, ENS, UPMC, CNRS; 24 rue Lhomond, 75005 Paris, France.}
\affil[4]{Institut Universitaire de France}

\leadauthor{Vernet}

\significancestatement{The Seebeck effect is the conversion of heat into electricity, usually achieved by thermoelectric devices using solid electrical conductors or semiconductors. Here is reported the first evidence of this effect at the interface between two metals that are liquid at room temperature, gallium and mercury. The liquid nature of the interface significantly alters the usual temperature distribution, leading to an abnormally high current density near the boundaries. In the bulk, the thermoelectric current interacts with a magnetic field to produce efficient thermoelectric pumping of fluids. This effect may be of prime importance in several industrial and astrophysical systems, such as the promising liquid-metal batteries and Jupiter's magnetic field.}

\equalauthors{\textsuperscript{1}M.V.(Author One) contributed equally to this work with S.F. (Author Two) and C.G. (Author Three).}
\correspondingauthor{\textsuperscript{2}To whom correspondence should be addressed. E-mail: christophe.gissinger@phys.ens.fr}

\keywords{Keyword 1 $|$ Keyword 2 $|$ Keyword 3 $|$ ...}

\begin{abstract}
We present experimental evidence of a thermoelectric effect at the interface between two liquid metals. Using superimposed layers of mercury and gallium in a cylindrical vessel operating at room temperature, we provide a direct measurement of the electric current generated by the presence of a thermal gradient along a liquid-liquid interface. At the interface between two liquids, temperature gradients induced by thermal convection lead to a complex geometry of electric currents, ultimately generating current densities near boundaries that are significantly higher than those observed in conventional solid-state thermoelectricity. When a magnetic field is applied to the experiment, an azimuthal shear flow, exhibiting opposite circulation in each layer, is generated. Depending on the value of the magnetic field, two different flow regimes are identified, in good agreement with a model based on the spatial distribution of thermoelectric currents, which has no equivalent in solid systems. Finally, we discuss various applications of this new effect, such as the efficiency of liquid metal batteries. \\
(published article available at https://www.pnas.org/doi/abs/10.1073/pnas.2320704121)
\end{abstract}

\dates{This manuscript was compiled on \today}
\doi{\url{www.pnas.org/cgi/doi/10.1073/pnas.XXXXXXXXXX}}

\maketitle
\thispagestyle{firststyle}
\ifthenelse{\boolean{shortarticle}}{\ifthenelse{\boolean{singlecolumn}}{\abscontentformatted}{\abscontent}}{}

\firstpage{4}

\dropcap{T}hermoelectricity describes the conversion of heat into electricity and vice versa. This captivating interplay has long intrigued physicists, as it offers a glimpse into the complex relationship between energy, temperature and matter~\cite{goldsmid2010introduction}. 

%
\tcr{ The thermoelectric Seebeck effect is perhaps the best illustration of this: when a temperature gradient is established at the junction of two electrically conducting materials, a thermoelectric current flows between the "hot" and "cold" regions. This configuration can be achieved very simply by layering two metals atop each other and applying a horizontal temperature gradient along the interface.}

In addition to its implications for fundamental physics, thermoelectricity has left an indelible mark on modern engineering thanks to the many applications developed over the last century. For example, thermocouples are widely used as temperature sensors, while emerging applications include thermoelectric coolers for portable refrigeration~\cite{zhao2014review}, or the use of thermoelectric materials in space missions for their ability to generate electricity from temperature differences in harsh environments~\cite{yang2006thermoelectric}. \tcr{Thermoelectricity is an environmentally friendly technology for converting waste heat into electrical energy.}

Thermoelectricity also extends to liquid systems, such as electrolytes~\cite{gunawan2013liquid} , liquid metals, or semi-conductors. During the growth of a semiconductor crystal~\cite{garandet1999bridgman} or the solidification of a metal alloy~\cite{moreau1993thermoelectric}, a thermoelectric current naturally appears at the liquid-solid interface due to the Seebeck effect. When subjected to a magnetic field, these currents can then produce significant flow motions in the melt.  This surprising effect traces back to the pioneering work of Shercliff~\cite{shercliff_1979a,Shercliff_1979b}, who introduced the concept of thermoelectric magnetohydrodynamics (TEMHD) to describe the interaction between a liquid metal and the container wall: when a magnetic field ${\bf B}$ and a temperature gradient are applied to a solid-liquid interface, the thermoelectric current ${\bf J}$ generated by the Seebeck effect interacts with the magnetic field to produce a Lorentz force ${\bf J\times B}$, which drives significant flow motions. Since Shercliff, only a few studies have provided experimental data on this effect. In the context of fusion energy, where TEMHD-induced flows can provide an effective cooling blanket~\cite{kaita2007_Li_stability,tokamak_li_blanket_2018,compass_U_2021}, a single experiment has reported velocity measurements in a divertor made of liquid lithium~\cite{jaworski2010_PRL_thermoE} heated by an electron beam. More recently, temperature measurements have been reported in an experiment that suggests an interesting interaction between thermoelectricity and magneto-convection producing periodic oscillations~\cite{Aurnou_2022}.

This paper reports the first experimental evidence of thermoelectricity at the interface between two liquid layers. This configuration is different from the classical thermoelectric effect, as the vessel walls, electrically insulating, are not involved in the generation of the current, which now occurs along a free interface between the two fluids. In particular, the temperature and current density distributions are different from the classical situation. The interest of our study is twofold. First, by using two liquid metals at room temperature, we aim to provide quantitative measurements of velocity, temperature, and electric potential associated with a simple theoretical model to describe precisely the dynamics of this new type of thermoelectricity. Second, these experimental results can be extrapolated to make predictions for several industrial and astrophysical systems where this effect can play a major role, in particular liquid metal batteries and Jupiter's magnetic field.\\

\section*{Experimental setup}

\begin{figure}[h!]
    \centering
    \includegraphics[scale=0.25]{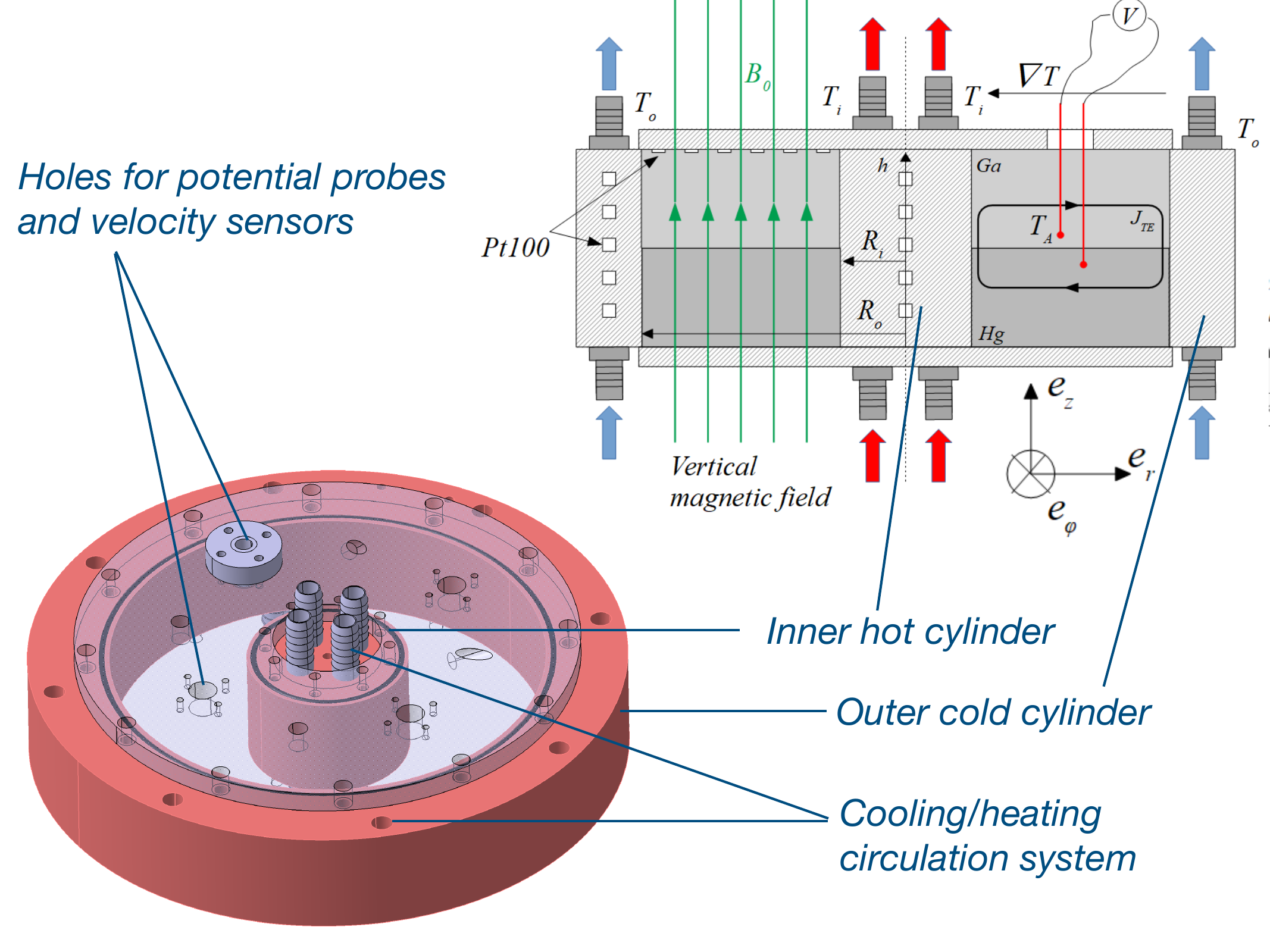}
    \caption{Sketch of the experiment. A cylindrical vessel made of two concentric, electrically insulating cylinders with radii $R_i=37$mm and $R_o=100$mm and height $50$mm is filled with half mercury, half gallium, forming a liquid metal interface. All boundaries are electrically insulating, ensuring complete electrical insulation of the two liquid metals from the outside world. The fluids are subjected to a thermal gradient due to a temperature difference between the two cylinders $\Delta T_0=T_i-T_o$. Thermoelectric potential and flow velocities are measured in the middle of the gap (see text). A vertical magnetic field up to $80$mT can be applied to the experiment. $J_{TE}$ represents a simplified distribution of thermoelectric currents, but only in the limit of very low thermal gradients or solidified metals (see text).} 
   \label{fig:schema_GALLIMERT}
\end{figure}

The experiment consists of a cylindrical annulus with a rectangular cross-section. The height is $h=50~mm$, and the radii of the inner and outer cylinders are respectively $R_i=37~mm$ and $R_o=100~mm$, corresponding to an aspect ratio close to one $\Gamma = L/h \sim 1.26$ with $L=(R_o-R_i)$ the cylindrical gap. (see Fig.\ref{fig:schema_GALLIMERT}).
The tank is filled with a layer of liquid gallium on top of an equally thick layer of liquid mercury. To avoid solidification of the gallium, which has a melting point of $29.7~^\circ C$, the tank is maintained at $35~^\circ C$ at least. To our knowledge, this is the first experiment on the dynamics of a gallium-mercury interface, providing a direct study of a conducting \tcr{liquid-liquid} interface at room temperature, mercury and gallium are almost immiscible. To maintain the immiscibility of the two fluids, all our experiments are limited to $T<80~^\circ C$. 

\tcr{To avoid mixing the two layers, the mercury is first introduced into the tank. The liquid gallium is then gently deposited on the surface of the mercury through a tube in which the flow is kept at a very low rate. The binary Hg/Ga phase diagram confirms the proper separation of the two liquid metals:}
 at this temperature, the mercury layer contains $3\%$ mass gallium at most, and \tcr{the interface remains well defined}~\cite{guminski1993ga}. The inner and outer cylinders are made of copper and electrically insulated from fluids by an epoxy resin \textit{Duralco 128}. The endcaps are $10~mm$ thick, electrically insulating PEEK plates. Both cylinders are connected to thermal baths to impose a radial temperature gradient.  The inner cylinder is heated by water circulation controlled by a refrigeration circulator \textit{Lauda 1845}  and the heat is removed from the outer cylinder by an oil circulation system controlled by a \textit{Lauda T10000} thermal bath. Some of our results are obtained in the presence of a magnetic field. For this purpose, the tank is placed between two large Helmholtz coils with an inner diameter of 500 mm powered by a DC current supply \textit{ITECH IT6015D 80~V-450~A}, which produces a constant and homogeneous vertical magnetic field of $80~mT$ maximum. The experiment can thus be controlled by two external parameters, namely the applied magnetic field $B_0$ and the temperature difference $\Delta T_0=T_i-T_o$ imposed between the two cylinders.

Temperature is measured \tcr{inside the} inner and outer cylinders, and in the tank, using Pt100 platinum resistance sensors. Five sensors are evenly distributed along a vertical line in each cylinder, while $14$ sensors are glued to the top endcap,  in contact with the gallium, along a line running from the inner to the outer cylinder (labeled $2$ to $15$ in the following). Four holes are drilled in the top endcap for various measurements: flow velocity and thermoelectric currents are obtained using electric potential measurements, while Hall probes are used to measure the magnetic field.  Temperature measurements are acquired using a \textit{Keithley 3706A signal-switching multimeter}, while potential measurements, particularly weak, are processed using a nano-voltmeter (Keysight 34420A).  All signals are then transmitted to the computer via a data acquisition card \textit{National Instrument 6212} controlled by scripts \textit{Python}.\\

\begin{figure}[h!]
    \centering
    \includegraphics[scale=0.25]{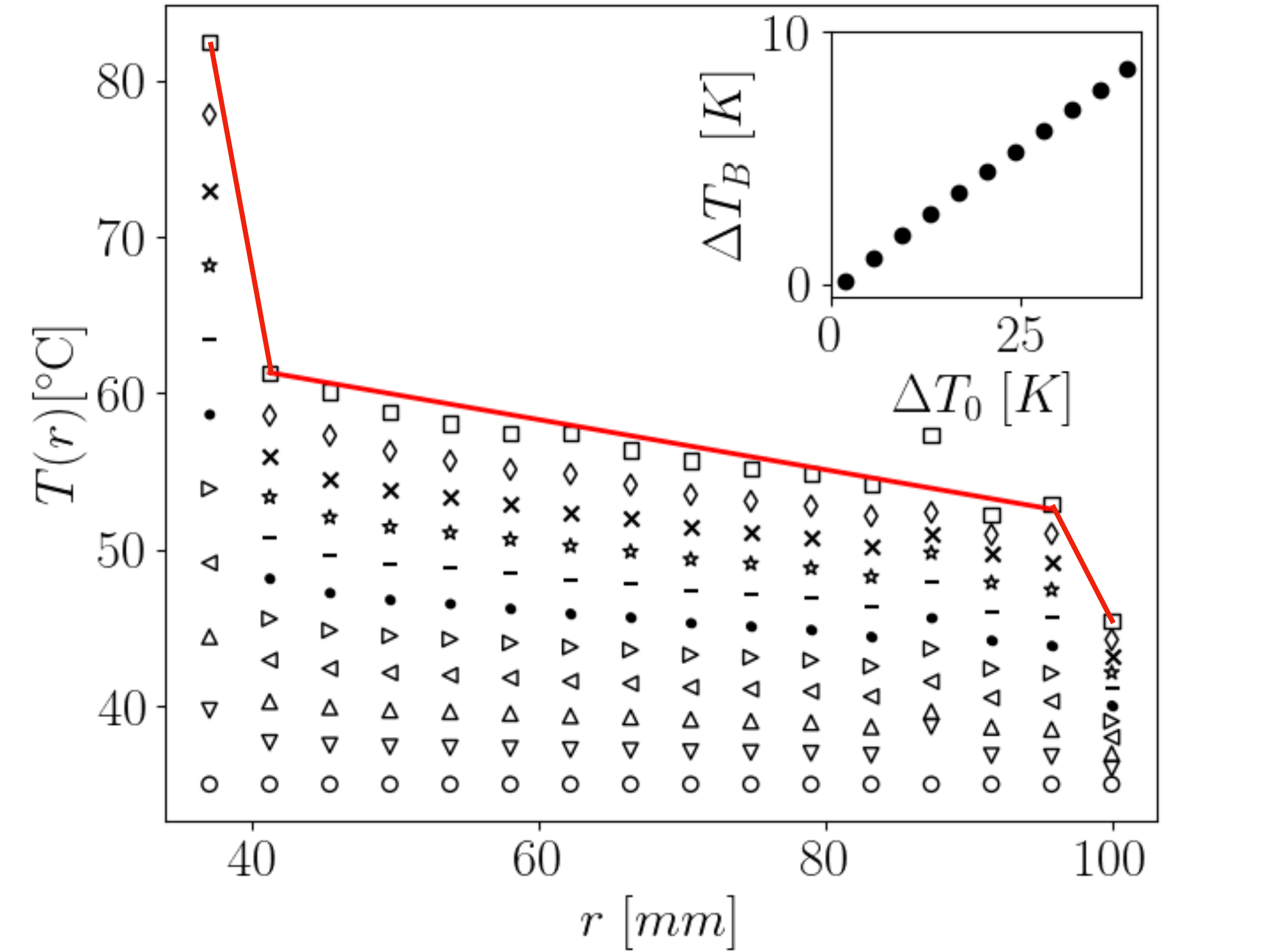}   
    \caption{ Radial temperature profile (measured at the top endcap) for different applied temperature differences $\Delta T_0=T_i-T_o$, for $B_0=0$. Temperatures at the first and last radial positions are measured inside the cylinders. Inset: Time-averaged temperature difference $\Delta T_{B}$ as a function of $\Delta T_0$, where $\Delta T_B=T_{15}-T_2$ is obtained using temperatures measured in gallium, at $5$ mm from the cylinders. Legend: $\Delta T_0=\text{0K}$ $(\circ)$, $\Delta T_0=\text{4K}$ $(\nabla)$, $\Delta T_0=\text{7K}$ ($\triangle$), $\Delta T_0=\text{11K}$ $(\triangleleft)$, $\Delta T_0=\text{15K}$ $(\triangleright)$, $\Delta T_0=\text{18K}$ $(\cdot)$, $\Delta T_0=\text{22K}$ $(-)$, $\Delta T_0=\text{26K}$ $(\star)$, $\Delta T_0=\text{30K}$ $(\times)$, $\Delta T_0=\text{33K}$ $(\Diamond)$, $\Delta T_0=\text{37K}$ $(\Box)$. The red curve is a linear fit of the piece-wise linear temperature profile in the case $\Delta T_0=\text{37K}$.}
    \label{fig:Temperature_profile}
\end{figure}

With two liquid layers, the temperature distribution responsible for the thermoelectric effect is entirely governed by fluid motions on either side of the interface. Indeed, the temperature gradient between the cylinders generates horizontal thermal convection in both layers, with typical Rayleigh numbers \tcr{of the order of $Ra=[10^4-10^5]$ (See SI Appendix for calculation), where $Ra = \alpha g \Delta T_0 \Delta R^3/\kappa\nu$ and $\alpha$ is the thermal dilatation coefficient, $\kappa$ is the thermal diffusivity and $\nu$ is the kinematic viscosity. } 
\tcr{For the Rayleigh numbers reported here, vigorous convection is expected. Although determination of the exact regime would require a separate study, it is plausible that our intermediate values of $Ra$ favor boundary-layer-dominated heat transfer, characterized by efficient turbulent heat transport in the bulk, and significant diffusive transport in the thin thermal boundary layers. This interpretation is confirmed by our temperature measurements:}

Fig.\ref{fig:Temperature_profile}  shows the temperature profile measured in the gallium layer, at the top endcaps, for a series of runs at $B_0=0$ and $\Delta T_0$ ranging from $0$ to $37$ K. It shows that the convective motions, although weak, are sufficient to transport heat and significantly flatten the temperature profile in the bulk. \tcr{This scenario markedly contrasts with the typical diffusive thermal gradient observed in solids}.
 Most of the temperature drop is therefore confined to thin thermal boundary layers close to the cylinders. The inset in Fig.~ \ref{fig:Temperature_profile} shows, however, that the temperature gradient in the volume \tcr{$\Delta T_B=T_{15}-T_2$} depends linearly on the applied temperature drop $\Delta T_0$.  As liquid metals are very good thermal conductors, we expect the interface temperature to follow this profile closely.

\section*{Seebeck effect}

\tcr{In each fluid layer, the Ohm's law in the presence of a thermal gradient reads}:

\begin{align}
    \frac{\bm{j}}{\sigma} = \bm{E} - S\bm{\nabla}T,
    \label{eq:Ohm_law}
\end{align}
where $\bm{j}$ is the electric current density, $\bm{E}$ is the electric field, $T$ is the temperature, $\sigma$ is the electrical conductivity and $S$ is the Seebeck coefficient. \tcr{For gallium and mercury, the values are given as $\sigma_{Ga} = 3.87\times 10^6~S.m^{-1}$, $\sigma_{Hg} = 1.1\times 10^6~S.m^{-1}$,  $S_{Hg}=-6.5~\mu V.K^{-1}$ and $S_{Ga}=0.5.~\mu V.K^{-1}$~\cite{Bradley63}.} 

\tcr{The production of thermoelectric current is made possible because the Seebeck coefficient $S$ depends not only on temperature but also  the substance:} in a uniform medium, the electric field is rearranged to compensate for the Seebeck effect and prevent the emergence of an electric current, $\bm{E}=-S\bm{\nabla}T$, a consequence of the fact that $\bm{\nabla}\times(S\bm{\nabla}T)=0$. To generate a net thermoelectric current, it is therefore necessary to misalign the temperature and Seebeck coefficient gradients, which can be achieved simply by generating a thermal gradient along an interface between two metals. 
\tcr{In the quasi-static limit,  $\bm{\nabla}\times\bm{E}=0$  allows us to write $\bm{E}=-\bm{\nabla}V$. 
In addition,  charge conservation $\bm{\nabla}\cdot\bm{j}=0$ implies that the electric potential follows a Poisson equation in each layer:}
\begin{align}
 \nabla^2 V = -S\nabla^2 T 
\label{Poisson}
\end{align}
\tcr{Combined with the appropriate boundary conditions at the interface between the two metals, these equations describe the generation of a Seebeck effect between the Gallium and Mercury layers. The detailed solution of equation (\ref{Poisson}) provides $V$, $\bm{j}$, and the corresponding magnetic field $B$. It is tedious enough to have been left in the Supp. Mat. and simplified by using cartesian geometry and a temperature field independent of $z$. This simplified model shows that an electric current can flow through liquid metals in response to a horizontal thermal gradient, even with the unusual geometry involving complete short-circuiting of the two layers along the interface. More precisely, the thermoelectric current depends critically on the temperature profile at the interface and it exhibits a linear dependence on the effective conductivity, $\tilde{\sigma} = \sigma_{Hg} \sigma_{Ga}/(\sigma_{Hg}+\sigma_{Ga})$ and the difference in Seebeck coefficients, $\Delta S = S_{Hg} - S_{Ga}$. In addition, calculations show that the thermoelectric current loop induces a measurable voltage drop between mercury and gallium.
}

\begin{figure}[h!]
    \centering
    \includegraphics[scale=0.6]{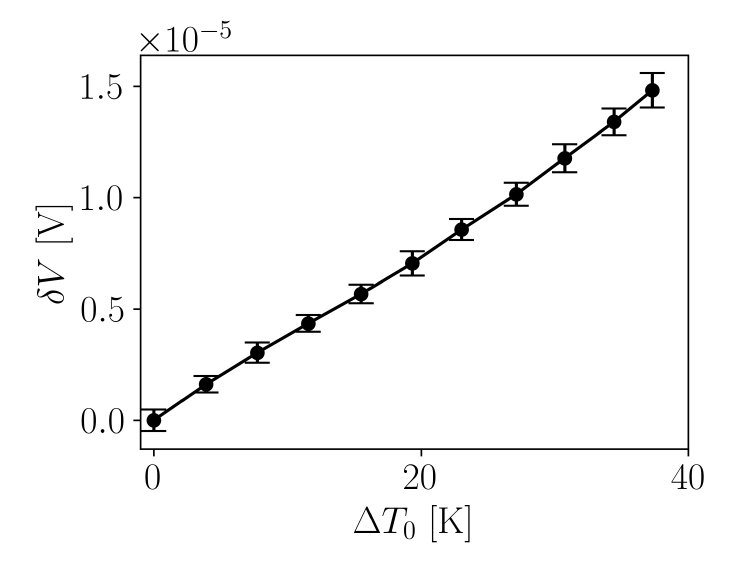}
    \caption{Thermoelectric potential as a function of the applied temperature difference $\Delta T_0$, for $B_0=0$. The error bars correspond to the standard deviation of the time signal of the electric potential \tcr{and therefore reflect a certain degree of unsteadiness induced by turbulent convection.}}
    \label{fig:pot_TE}
\end{figure}

Experimentally, the thermoelectric effect can be evaluated directly via the electric potential difference between two points on either side of the liquid-metal interface (see Fig. ~\ref{fig:schema_GALLIMERT}), \tcr{related to the current by}:
\tcr{
\begin{align}
    \delta V = -\int_A^B\frac{\bm{j}}{\sigma} \cdot\bm{dl} -\int_A^BS\bm{\nabla}T\cdot\bm{dl} 
    \label{dphi}
\end{align}
}
where \tcr{this} integration \tcr{of equation (\ref{eq:Ohm_law})} can be done along any path from $A$ to $B$. In the experiment, we measure this voltage between two nickel wires, fully coated except at their ends, and placed so that the wire tips are located at mid-radius $r=r_i+L/2$, inside each layer, at approximately  3 mm from the interface. 
\tcr{Fig.~\ref{fig:pot_TE}(a)} shows the evolution of voltage as a function of the imposed temperature gradient $\Delta T_0$. The measured voltage displays a linear evolution with  $\Delta T_0$ and reaches about $15~\mu V$ for $\Delta T_0\sim 37$K, therefore demonstrating the existence of a thermoelectric effect generated at the interface between two liquid metals. 
\tcr{ 
In agreement with the theoretical predictions of our simplified model, the voltage $\delta V$ is approximately linearly related to the temperature difference applied between the two cylinders. However, accurately determining the maximum voltage measured in the experiment is challenging due to several factors not accounted for in the theory. These include geometric effects, contact properties at the interface, oxidation of gallium, miscibility thickness, convective motions, and the vertical thermal gradient. Each of these factors can significantly influence the numerical value of $\delta V$.}\\

\tcr{
In fact, the liquid nature of the two layers is the key to understanding the magnitude of this thermoelectric effect. Unlike solid-state thermoelectricity and thermocouples, which involve connected electrical wires, the geometry of currents in this case is not prescribed, and thermoelectric currents are subject to the powerful convective motions of liquids. In the next section, we will show how turbulent convection, by modifying the temperature profile along the interface, leads to a complex distribution of thermoelectric currents in the bulk flow and particularly high current densities near thermal boundary layers.}

\section*{Geometry of the electric currents}

\begin{SCfigure*}
\centering 
  \includegraphics[width=11.4cm]{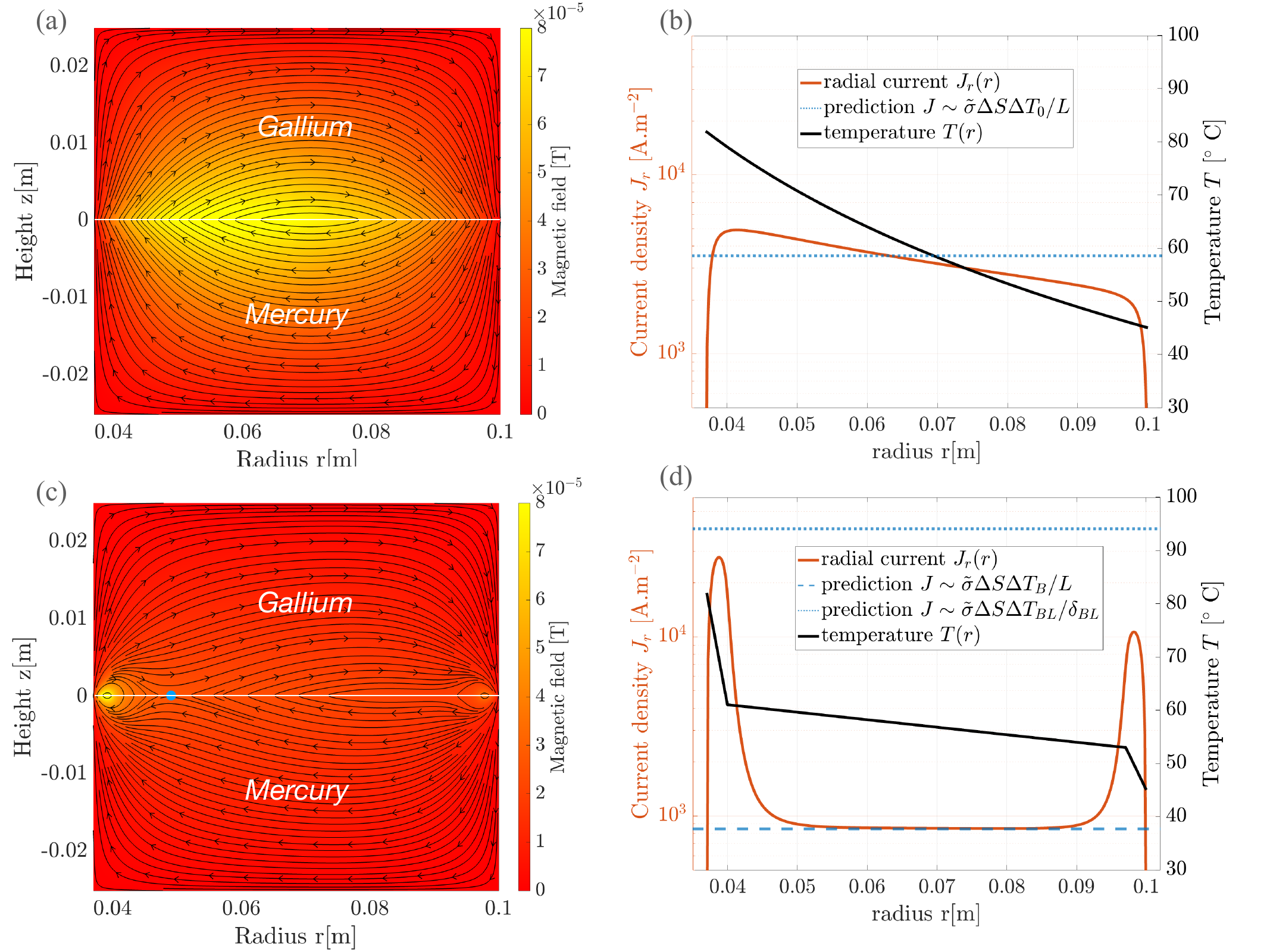}
\caption{Numerical integration of equation (\ref{eq:Ohm_law}) using the parameters of the experimental setup (see Method section) and a piecewise linear thermal gradient, for $B_0=0$. 
(a) colorplot of the induced magnetic field and associated current streamlines, in the case of a purely conductive temperature solution. (b) radial profiles of the corresponding temperature (black) and the  radial current induced at $z=1$ mm from the interface (red). (c) and (d) are the same, but 
for a piecewise temperature gradient typical of convection. Near the cylinders, the thermal boundary layers generate a very large current density, 10 times larger than the value expected with solid-state conventional thermoelectricity. The dashed (resp. dashed-dotted) line shows the simple prediction (\ref{predictionJ}) for bulk (resp. boundary) density currents.}
\label{fig:simu}
\end{SCfigure*}

\tcr{This complex dependence on the temperature profile} contrasts sharply with what is observed in solid-state thermoelectricity, and even in classical thermoelectric MHD, where the two temperatures imposed at the conducting walls always drive the current measured in the bulk.  This is because the temperature profile is extremely different from the linear thermal gradient observed in solid conductors, and the geometry of the current becomes different from the naive picture described above and sketched in  Fig.\ref{fig:schema_GALLIMERT}. To understand how a \tcr{liquid-liquid} interface affects the distribution of thermoelectric currents, we carried out 2D axisymmetric numerical simulations of Ohm's relation (\ref{eq:Ohm_law}) in the cylindrical geometry of the experiment and using the physical properties of gallium and mercury (see the Method section).

Although only the numerical integration is discussed here, the Supplementary Materials show that identical results are obtained with the analytical calculation (see Supp. Mat. for a detailed description of the analytical model).  
Fig.~\ref{fig:simu}(a) shows a simulation computed using boundary temperatures obtained experimentally at $\Delta T_0=37K$ (namely  $T_h=82^\circ C$ and $T_c=45^\circ C$ ) but with a temperature profile $T=A\log(r)+B$, solution of $\nabla^2T=0$, as if the metals were solid. In this case, the field geometry is as expected, with an electric current predominantly horizontal at the center of the cell, forming a poloidal loop around the interface.

\tcr{The order of magnitude of the bulk current can be simply recovered by  performing the curvilinear integral along a closed loop $\mathcal{C}$ of equation (\ref{eq:Ohm_law}), which leads to $\oint_\mathcal{C}\bm{j}\cdot \bm{dl}/\sigma \approx -\Delta S \Delta T$
with $\Delta S$ assumed independent of $T$, and $\Delta T$ is the temperature difference between the two points where $\mathcal{C}$ crosses the interface. 
 By assuming a predominantly horizontal current density in the bulk, away from the boundaries, so that charge conservation leads to an identical horizontal current $|j|$ in each layer (ignoring curvature),  this relation  can be integrated and provides a simple estimate of the current density: 
\begin{equation}
 j\sim \frac{\Delta S\Delta T}{\ell} \tilde{\sigma} 
\label{predictionJ}
\end{equation}
where $\Delta T=T_h-T_c$ is the temperature difference driving the currents with $T_h$ (resp. $T_c$) representing the hot (resp. cold) temperature and $\ell$ is the typical length of temperature variation responsible for the thermoelectric current. As usual, the amplitude of the thermoelectric current thus depends on the jump of Seebeck coefficients between the two materials and the temperature difference between the "hot" and "cold" regions of the interface. In the case of solid metals, it is clear that $\ell=L$ and $T_h-T_c=\Delta T_0$, and Fig.~\ref{fig:simu}(b) shows that the radial current in the middle of the gap is of the order of $J\sim \tilde{\sigma}\Delta S\Delta T_0/L$ (blue dotted line), as expected in solid-state thermoelectricity.}\\
But as shown in Fig.\ref{fig:Temperature_profile}, the actual temperature profile for liquid metals is radically different and instead displays a piecewise constant gradient involving two very strong thermal gradients confined to thin boundary layers of thickness $\delta_{BL}$, connected by a gentler linear variation in the bulk. Such a profile is forbidden in the presence of a solid boundary and is only possible here due to vigorous thermal convection in the two liquids on either side of the interface. \tcr{In the presence of liquid layers, the choice of  $T_h$, $T_c$, and $\ell$ is thus highly nontrivial}. Fig.~\ref{fig:simu}(c) shows a typical numerical integration using such an experimental profile (i.e. the piecewise linear fit shown in red in Fig.\ref{fig:Temperature_profile}). Far from the boundaries, the geometry of currents remains relatively similar to the previous case. The corresponding radial profile in Fig. \ref{fig:simu}(b) shows that currents reach a plateau in the bulk, with a magnitude that corresponds exactly to the prediction $J\sim \tilde{\sigma}\Delta S\Delta T_B/L$ (dashed line). These simulations therefore show that the thermoelectrical current generated in the bulk is not directly due to the temperature drop $\Delta T_0$ imposed at the boundaries but is rather driven by the lower thermal gradient that subsequently occurs in the bulk outside the boundary layers, characterized by the temperature difference $\Delta T_B$.

On the other hand, Fig.\ref{fig:simu}(c) clearly shows that two additional thermoelectric current loops are induced by the large temperature gradient in the thermal boundary layers. These currents are located fairly close to the cylinders, but the current density is surprisingly high: for $\Delta T_0=37K$, it can reach $j\sim 3\times 10^4 A/m^2$  (see Fig.\ref{fig:simu},d), $40$ times higher than in the bulk. Interestingly, this value is also one order of magnitude higher than the one expected in the case of solid metals (Fig.\ref{fig:simu},b). This high value can easily be understood as a local generation of thermoelectric currents by the strong temperature gradient $\Delta T_{BL}$ in the thermal boundary layer of thickness $\delta_{BL}$. Hence, the estimate $j\sim\tilde{\sigma}\Delta T_{BL}\Delta S/\delta_{BL}$, where $\Delta T_{BL}$ is the temperature drop inside the boundary layer provides the correct value of this anomalously high density current (dotted line in Fig.\ref{fig:simu}(d)).  
The liquid nature of the interface therefore produces a non-trivial distribution of thermoelectric currents, well illustrated by the saddle point formed by the currents at the interface (indicated by the blue point in Fig.\ref{fig:simu}(c) ). The radial position of this saddle point depends on the details of the configuration, but its existence is an unavoidable consequence of the non-linear temperature gradient produced in the liquids. 

These high current densities cannot be directly detected in the experiment due to their confinement near the walls, where electrical measurements are unavailable. \tcr{However, in the next section, we demonstrate that surface velocity measurements, conducted in the presence of a magnetic field applied to the layers, can infer the existence of these high current densities and provide an accurate estimate of the value of bulk currents.} Note that the analytical calculation in Supp. Mat shows that this peculiar geometry of the currents is driven by the temperature at the interface and can not be observed in the case of a liquid in contact with a conducting wall, for which the thermal gradient is constant at the liquid/solid boundary. This highlights the essential role of the fluid motions near the interface for the dynamics of thermoelectric currents. \\

\section*{Thermoelectric magnetohydrodynamics}

\begin{figure}[h!]
    \centering
    \includegraphics[scale=0.58]{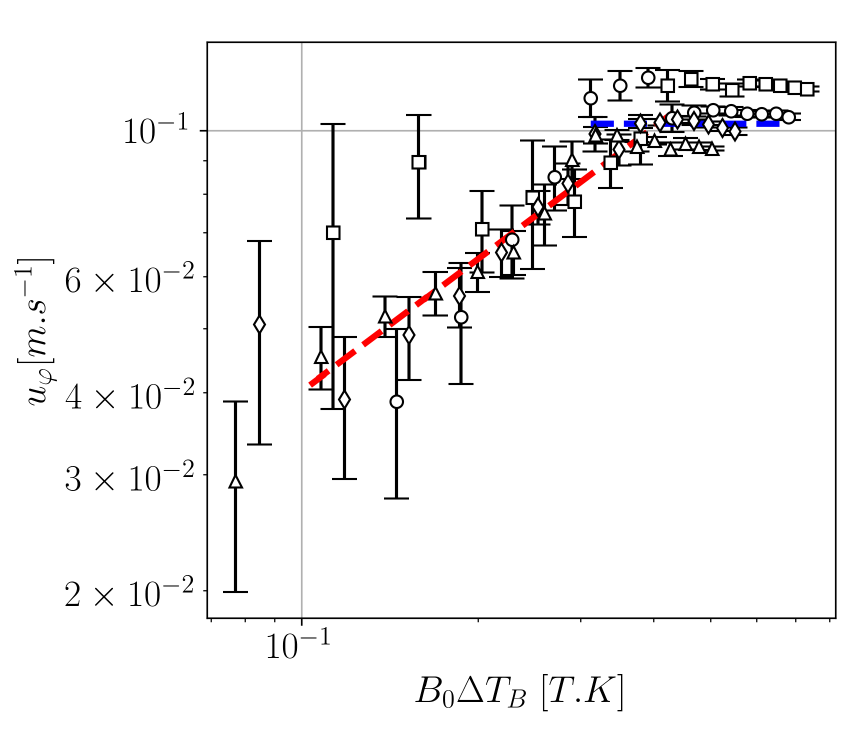}
    \caption{Time-averaged azimuthal velocity as a function of the product $B_0\Delta T_B~[T.K]$. $\Delta T_0=23$K  $(\triangle)$, $\Delta T_0=29$K  $(\Diamond)$, $\Delta T_0=33$K  $(\Box)$, $\Delta T_0=37$K  $(\circ)$. The error bars correspond to the standard deviation of the velocity. Two different regimes are observed, that can be relatively well fitted by our predictions (\ref{eq:u_23}), red dashed line and (\ref{eq:u_hartmann_TE}), blue dashed line.}
    \label{fig:U_TE_vs_BdT}
\end{figure}

The experiment is now subjected to a vertical homogeneous magnetic field $B_0$ using the two coils. \tcr{In the presence of a magnetic field, Ohm's law (\ref{eq:Ohm_law}) is modified as follows to take into account the magnetic induction:
\begin{align}
	\frac{\bm{j}}{\sigma} = -\bm{\nabla}V + \bm{u}\times\bm{B} - S\bm{\nabla}T,
\end{align}
where $\bm{u}$ denotes the velocity field and $\bm{B}$ is the magnetic field.} In the presence of this field, the horizontal thermoelectric currents described above generate an azimuthal Lorentz force, directly proportional to the product of $B_0$ and the temperature difference $\Delta T_B$ producing the currents. In this configuration, the azimuthal velocity $u_\varphi$ can be obtained by measuring the voltage between two wires both located in liquid gallium (12 mm above the mercury-gallium interface), so that the contribution of the thermoelectric current can be neglected~\cite{molokov2007velocity}. In Fig.\ref{fig:U_TE_vs_BdT}, we report the time-averaged value of $u_\varphi$ as a function of $B_0$, for different fixed values of the temperature difference $\Delta T_0$. Even a moderate temperature gradient can produce a relatively vigorous motion of the liquid gallium, which reaches nearly $\sim 15 $cm/s for $B_0=56$mT and $\Delta T_0=37K$.  Note that, as the current changes sign in each layer, this Lorentz force causes the two liquid metals to rotate in opposite directions, generating a strong azimuthal shear flow at the interface.  In what follows, we only measure the velocity field generated in the upper layer of liquid gallium, but it should be kept in mind that a similar flow occurs in the bottom layer (albeit somewhat weaker due to the lower conductivity and higher density of mercury). If the applied magnetic field changes sign, the direction of the azimuthal velocity is reversed, as expected.
The flow has two distinct behaviors, depending on the relative magnitudes of the magnetic and velocity fields. At a small magnetic field, as long as $u_\varphi<10$cm/s or so, the velocity increases rapidly with the magnetic field, and most of the data collapse to the prediction $u_\varphi\propto(B_0)^{2/3}$. This exponent has been reported in several recent experimental and numerical studies, in which a conducting fluid is driven by an electromagnetic force ~\cite{Poye20,vernet21,Davidson22}. It is relatively simple to extend these previous studies to thermoelectric currents generated in the liquid gallium: as suggested by Fig.\ref{fig:simu}, the current density in the bulk is distributed over the entire layer $h/2$, so that the azimuthal Lorentz force balances the inertia $j_{TE}B_0\sim \rho u_ru_\varphi/r$. Near the endcap and the interface, the imbalance between the pressure gradient and vanishing centrifugal force produces a radial flow $u_r^{BL}$ in the viscous boundary layers, such that $u_\varphi^2/r \sim \nu u_r^{BL}/\delta_B^2$ with $\delta_B=\sqrt{\nu r/u_\varphi}$ the thickness of the Bödewadt boundary layer. Combining these two relations and using an incompressibility condition $2u_r^{BL}\delta_B\sim u_rh/2$,  we finally obtain a prediction for the mean azimuthal velocity field:

\begin{align}
    u_\varphi \sim \left(\frac{j_{TE}(r)B_0h\sqrt{r}}{4\rho\sqrt{\nu}}\right)^{2/3}\sim  \left(\frac{\tilde{\sigma}\Delta S\Delta T_BB_0h\sqrt{r}}{4L\rho\sqrt{\nu}}\right)^{2/3}
\label{eq:u_23}
\end{align}
where we used $j_{TE}\sim \tilde{\sigma}\Delta S\Delta T_B/L$  to obtain the final expression. This prediction is indicated by the red dashed line in Fig.\ref{fig:U_TE_vs_BdT}. It shows reasonable agreement with the experiment, despite some scatter in the data. More importantly, this agreement confirms that the bulk temperature drop  $\Delta T_B$ (and not $\Delta T_0$) is responsible for driving the flow, at least in the middle of the gap.

At a sufficiently large magnetic field, the velocity field reaches a plateau, in which the flow no longer depends on the magnetic field and is driven solely by the temperature gradient at the interface. This regime is also relatively similar to what has been described for strongly magnetized flows subjected to external currents~\cite{Poye20,vernet21,Davidson22}. We briefly recall below the main derivation for this classical prediction, adapting it to the thermoelectric case. This plateau can be interpreted as a fully magnetized regime, in which the currents induced by the flow motions in the bulk become sufficiently large to oppose the applied thermoelectric currents, i.e. $\sigma u_\varphi B_0\sim j$. As a result, the thermoelectric currents flow through two thin Hartmann boundary layers generated at the endcap and at the interface (where the velocity must be zero due to the symmetry of the counter-rotating flow). The current density in these horizontal boundary layers can be estimated to $j\sim j_{TE}h/(4\delta_{Ha})$ where $\delta_{Ha}\sim \sqrt{\sigma/\rho\nu}/B_0$ is the thickness of Hartmann boundary layers. We then obtain a second prediction, independent of the magnetic field:
\begin{align}
    u_\varphi \sim \frac{j_{TE}}{4\sqrt{\rho\nu\sigma}} \sim \frac{\tilde{\sigma}\Delta S\Delta T_B}{4L\sqrt{\rho\nu\sigma}}
    \label{eq:u_hartmann_TE}
\end{align}
where again $j_{TE}\sim \tilde{\sigma}\Delta S \Delta T_B/L$ has been used. 
For $\Delta T_{B}\sim 8~K$ this prediction gives $u_\varphi \sim 13~cm.s^{-1}$ (blue dashed line in Fig.~\ref{fig:U_TE_vs_BdT}), which is in good agreement with the plateau measured at high magnetic field. 
\begin{figure}[h!]
    \centering
    \includegraphics[scale=0.25]{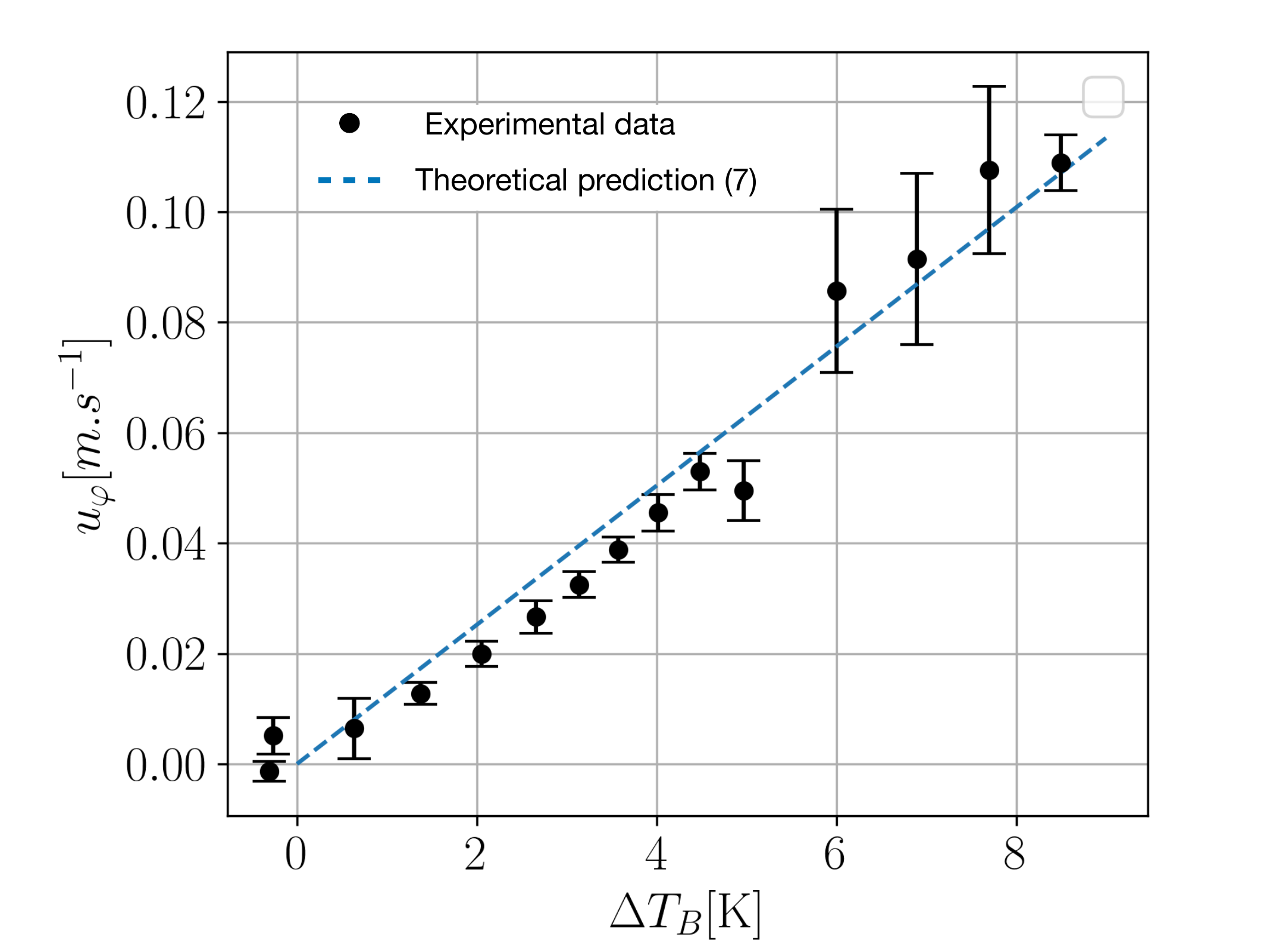} 
    \caption{Time-averaged velocity as a function of the bulk temperature difference $\Delta T_B~[K]$ for $B_0=56$mT (black circles), compared to prediction (\ref{eq:u_hartmann_TE}) (dashed line). Points above $\Delta T_B = 6$K were performed at constant imposed temperature difference in the domain where evolving $B_0$ let the velocity invariant. The error bar corresponds to the standard deviation of the velocity.} 
    \label{fig:Uphi_vs_dT_bulk}
\end{figure}
To further test this prediction, we report in Fig.\ref{fig:Uphi_vs_dT_bulk} the azimuthal velocity $u_\varphi$ as a function of the measured bulk temperature gradient, showing that the flow depends linearly on the thermal gradient generated in the bulk and follows closely prediction (\ref{eq:u_hartmann_TE}) (blue dashed line in Fig.\ref{fig:Uphi_vs_dT_bulk}). Finally, note that the transition between the inertial-resistive regime (\ref{eq:u_23}) and the fully magnetized regime (\ref{eq:u_hartmann_TE}) should occur when magnetic and rotational effects are in balance, i.e. when the Elsasser number $\Lambda = \sigma B_0^2/\rho\Omega$ is close to unity~\cite{davidson_Pot_note_Bodewadt_Ha,vernet21} \tcr{where $\Omega = u_\varphi/r$}. The intersection of the two predictions in Fig.\ref{fig:U_TE_vs_BdT} is obtained for $\Lambda_c\simeq 0.9$, in agreement with this picture. \\

To go beyond these local measurements and demonstrate the existence of large current densities at the boundaries, we carried out a few runs without the top endcap, so that the gallium phase displays a free interface. To prevent excessive oxidation of the gallium, the latter is in contact with a thin layer of hydrochloric acid HCl, which then replaces the endcap.  Using the presence of small oxides on the free surface, the velocity field is characterized by particle tracking using a CMOS camera with a resolution of $1080$x$2049$ and an acquisition frequency of $30$Hz. 
This approach has several drawbacks compared with local potential measurements: the density of the oxides is quite different from pure gallium, and their motion is slowed down by the friction from the HCl layer. This considerably \tcr{underestimates} the magnitude of the flow immediately below the free surface. But it also offers some advantages. To our knowledge, this is the first direct visualization of the thermoelectric pumping of a liquid metal (see the movie in supplementary materials), which allows us to study the spatial structure of the flow. 
\begin{figure}[h!]
    \centering
    \includegraphics[scale=0.24]{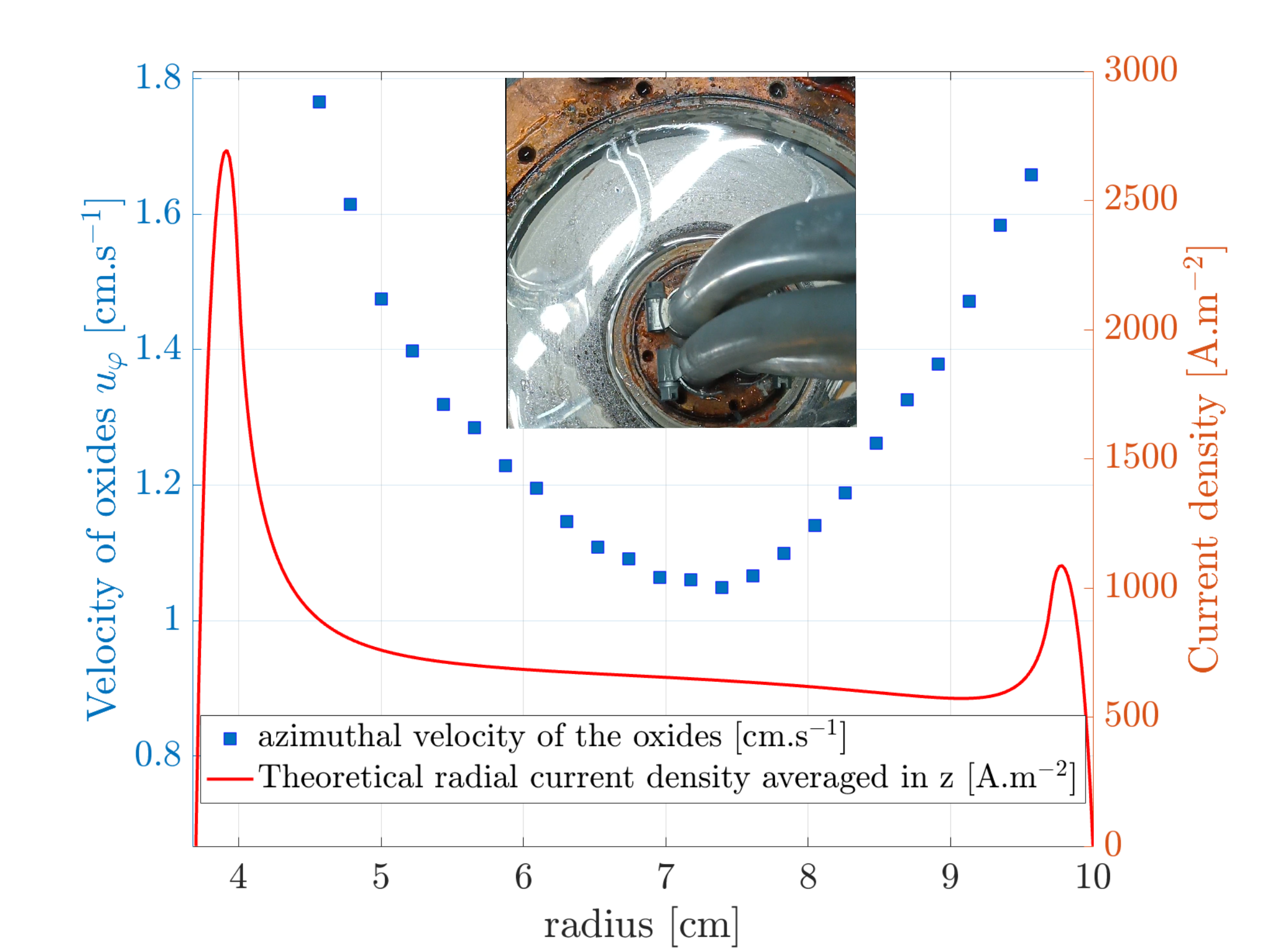}
    \caption{Radial profile of the azimuthal velocity $u_\varphi$ measured at the surface of the gallium for $B_0=36$mT and $\Delta T_0=37$K, when the top endcap is removed, using particle tracking of surface oxides. Near the outer cylinder, azimuthal velocity increases significantly with radius, due to the high current density generated at the boundaries.}
    \label{fig:movie3}
\end{figure}
Fig.~\ref{fig:movie3} shows the azimuthal velocity profile $u_\varphi$ obtained for $B_0=36$mT and $\Delta T_0=37$K.  At the surface, the measured velocity of the oxides is relatively fast, reaching $u_\varphi\sim 2 cm/s$ near the inner cylinder. Because of the drag produced by the $HCl$, it is difficult to deduce the absolute value of the velocity in the gallium phase immediately below this interface, but we expect the measured velocity profile to be a good proxy of the one in the bulk.  \tcr{Close to inner and outer radial boundaries, the azimuthal velocity $u_\varphi$ sharply {\it increases},} that can only be explained by the presence of an increasing magnetic forcing near the boundary. This additional rotation therefore provides an indirect measure of the large thermoelectric current density predicted by our calculations in Fig.\ref{fig:simu}. In Fig.\ref{fig:movie3}, we plot this theoretical profile of the radial current, averaged in $z$ over the whole layer of Gallium (red solid line). This current, induced by the thermal boundary layers, combines with the homogeneous magnetic field to produce a Lorentz force much larger at the boundaries. Although it is difficult to extrapolate from these measurements, it is interesting to note that the boundary current density, about $10$ times greater than that generated in the bulk, could lead to an azimuthal flow near the boundaries much faster than the one in the bulk. \\

\section*{Discussion and conclusion}
Although thermoelectric MHD has been discussed previously in the literature, the results reported here describe a different type of thermoelectricity. The liquid nature of the two conductors leads to a more complex temperature distribution, generating anomalously strong density currents near the boundaries and driving an azimuthal shear flow in the bulk. This situation can occur in a variety of contexts, and it is appropriate to conclude this paper with a brief discussion of these possible applications.

\begin{table}[h]
\centering
\caption{Main properties of the different components of a Liquid Metal Battery $Li||LiCl-KCl||Pb-Bi$~\cite{kollner2017_Rayleigh_Marangoni_LMB,van1955electrical_KCl-LiCl}}
\begin{tabular}{lrrr}
Species & Li & LiCl-KCl & Pb-Bi \\
\midrule
 Density $\rho~[kg.m^{-3}]$ & $484.7$ & $1597.9$ & $\sim 10^4$ \\
Viscosity $\nu~[m^2.s^{-1}]$ & $6.64\cdot 10^{-7}$ & $1.38\cdot 10^{-6}$ & $1.29\cdot 10^{-7}$ \\
Conductivity $\sigma~[S.m{-1}]$ & $3\cdot 10^6$ & $187.1$ & $7.85\cdot 10^5$ \\
\bottomrule
\end{tabular}
 \label{tab:LMB_type_Li}
\end{table}

Liquid metal batteries (LMBs) comprise three layers of different conducting fluids (top and bottom electrodes and a middle electrolyte) that self-segregate based on density and immiscibility and are subjected to electric current flowing through the fluids. Designed to store energy very efficiently, these low-cost, high-capacity, long-lasting, and easy-to-manufacture batteries could one day play a vital role in the massive expansion of renewable energy. 
Due to the high operating temperature of these systems, one could expect significant horizontal temperature gradients at the interfaces between liquid metals and the electrolyte. 
 A crude estimate can be made using the properties of lithium-bismuth batteries $Li|| LiCl-KCl|| Pb-Bi$, given in table \ref{tab:LMB_type_Li}~\cite{kollner2017_Rayleigh_Marangoni_LMB}. The Seebeck coefficient of liquid lithium is $S_{Li} = 26~\mu V.K^{-1}$~\cite{van1980thermoelectric_Li}. It is more difficult to estimate the Seebeck coefficient of the electrolyte, but values for $LiCl$ around $[100-1000]~\mu V.K^{-1}$ can be used here as an estimate of typical molten salt electrolytes. For a typical battery delivering $100 A$ and operating at $T>500~^\circ C$ during charging and discharging, the vertical magnetic field can be estimated at $1G$~\cite{Davidson22}. \tcr{For a typical cell with moderate size  $r\sim h \sim 20~cm$, applying a typical horizontal temperature gradient in the range $10-20 K$~\cite{kelley2018fluid} could produce thermoelectrical flows of  $u_\varphi\sim 3~mm.s^{-1}$ according to  prediction (\ref{eq:u_23}).} Such a flow magnitude is comparable to, perhaps larger than other phenomena expected in LMBs, such as Benard-Marangoni~\cite{kollner2017_Rayleigh_Marangoni_LMB} or flows induced by the Tayler instability~\cite{weber2014current}. Note that a similar flow in opposite direction is expected in the electrolyte layer. Unlike these other sources of motion, thermoelectric stirring does not rely on instability. With simple control of the horizontal thermal gradient in the cell, this shear flow could be used to significantly increase LMB efficiency by enabling the kinetic reaction and influencing the transfer of $Li^+$ ions through the electrolyte layer and into the Pb-Bi phase. 
 
\tcr{ Note, however, that these considerations are only valid in the absence of an externally imposed magnetic field. Such a field, often considered as a means of suppressing some undesirable instabilities, could then become harmful:  our flow predictions show that the Seebeck effect could produce a significant thermoelectric pumping, possibly capable of destabilizing the interface and thus short-circuiting the two electrodes.} 
 
The thermoelectric effect has also been proposed to explain some features of the magnetic fields of the Earth and Mercury~\cite{inglis1955theories,stevenson1987mercury}, where a thermoelectric interface is expected between liquid iron and semiconducting silicate rocks at the core-mantle boundary of these planets. The theoretical expressions reported here provide new quantitative predictions about the regimes eventually reached in these systems. Furthermore, the \tcr{liquid-liquid} interface specifically addressed here may be relevant to other astrophysical bodies. Jupiter is probably the best example. At $85\%$ of its radius, it exhibits an abrupt transition between an inner region of metallic hydrogen and an outer atmosphere of liquid molecular hydrogen. Since non-negligible meridional temperature variations are expected along this interface, it bears many similarities to the configuration described here. Here again, coefficients are relatively difficult to estimate, but let's assume that $\Delta S$ and $\tilde{\sigma}$ are both dominated by values of the semiconducting molecular hydrogen close to the transition with the metallic layer, such that $\Delta S\sim 1$~mV.K$^{-1}$ and $\tilde{\sigma}\sim 10^4$. 
\tcr{In this case, temperature variations of the order of  $1$~K would lead to a local azimuthal magnetic field $B_\varphi\sim \mu_0\tilde{\sigma}\Delta T\Delta S$ of the order of $10 \mu T$, a non-negligible fraction of the non-dipole radial magnetic field  reported recently~\cite{moore2018complex}. In addition, this thermoelectric current, presumably meridional, can interact with the planet's radial magnetic field to generate complex zonal flows.  Similar arguments could be made for stellar interiors at the transition between radiative and convective regions.}

A final comment must be made on the very large current density induced by thermal boundary layers. The \tcr{liquid-liquid} interface increases the current density by a factor of $L/\delta$ compared with a conventional solid thermocouple, where $L$ and $\delta$ represent the size of the thermocouple and the size of the thermal boundary layer respectively. In the context of a transition to sustainable energy sources, efficient waste heat recovery generally involves large-scale systems with a substantial temperature gradient, two ingredients that maximize $L/\delta$. In this case, using a liquid metal interface to convert heat into electricity may increase the efficiency of thermoelectric devices by several orders of magnitude. As the Prandtl number is small in liquid metals, the thermal layer is thicker than the viscous layer, which ensures that the boundary currents efficiently drive the fluids in the presence of a magnetic field. This possibility obviously requires further theoretical study, but it could offer an interesting new mechanism for converting heat into mechanical energy.

\matmethods{
\subsection*{Experimental measurements}

As shown in Fig.\ref{fig:schema_GALLIMERT}, the experiment is equipped with 4 holes on the top endcaps, located at $r=R_i+L/2$ through which various probes can be immersed in the liquid metals.  To measure the velocity field in the gallium layer, two nickel wires, completely insulated except at their conducting tips (noted $A$ and $B$ below) and separated by a distance $d=8$mm, are immersed in the liquid. The Seebeck coefficient of nickel is denoted $S_{Ni}$, and the electrical conductivity and Seebeck coefficient of the liquid metal are denoted $\sigma_{Ga}$ and $S_{Ga}$.
The electromotive force between points $A$ and $B$ is directly given by Ohm's law integrated over the distance between the wires:

\begin{equation}
e = \int_A^B \left( -S_{Ga}\nabla T + {\bf u\times B} - {\bf j}/\sigma_{Ga}\right)\cdot{\bf dl}
\end{equation}

By neglecting the induced currents, the voltage measured by the nano-voltmeter Keysight 34420A connected to the wires is:
\tcr{
\begin{equation}
e 
=(S_{Ni}-S_{Ga})(T_A-T_B) +UB_0d
\end{equation}
}
With $(S_{Ni}-S_{Ga})\sim 10 \mu $V.K$^{-1}$, the thermoelectric effect between the gallium and nickel wires introduces a velocity error $\delta U\sim (S_{Ni}-S_{Ga})(T_A-T_B)/(dB_0)$. For $B_0\sim 50$ mT and $(T_A-T_B)\sim [0.1-1]K$, this leads to $\delta U\sim2$ cm.s$^{-1}$ at most.
This offset is significantly smaller than our measured velocities and in practice has been systematically subtracted using the potential \tcr{$e(B_0=0)$} measured in the absence of magnetic field.\\

As explained in the main text, the measurement of the thermoelectric potential is based on the same technique, except that the two conducting tips are now located at different heights, so the tip of one of the wires is now immersed in the mercury layer. In this case, the magnetic field from the coils is zero, so $B_0$ reduces to the Earth's magnetic field. In this case, $uB_0d\sim10^{-8}$V, a value much smaller than the measured voltages, hence leading to the expression given in the main text.

\subsection*{Numerical modeling}

The equation (\ref{eq:Ohm_law}) has been numerically integrated in an axisymmetric cylindrical geometry using the same dimensions as the experiment and the physical properties of gallium and mercury. Specifically, we integrate the curl of the equation, so that it becomes a modified Poisson equation for the azimuthal magnetic field $B(r,z)$ :

\begin{equation}
\nabla^2 B=\frac{1}{\eta}\nabla S\times\nabla T - \partial_zB\frac{\partial_z\eta}{\eta}
\end{equation}
where $\eta=1/(\mu_0\sigma)$ is the magnetic diffusivity. This equation is solved by a Finite Difference Method using a 2$^{nd}$ order numerical scheme with the central difference in space. The magnetic field is set to zero at the boundaries to model an insulating vessel.
The interface between the two layers is modeled by taking $\eta(z)=\eta_{Hg}-(\eta_{Hg}-\eta_{Ga})(1+\tanh(z/z_i))/2$ and $S(z)=S_{Hg}-(S_{Hg}-S_{Ga})(1+\tanh(z/z_i))/2$ where $z_i$ is the typical thickness of the effective interface, taken as small as possible and fixed at $2$ mm in the results reported here. The temperature depends only on $r$ and is taken either as the conductive solution in cylindrical geometry $T(r)=A\ln r +B$ (using the same boundary temperatures $T(r_i)$ and $T(r_o)$ as the experimental temperatures measured in the cylinders) or as a piecewise constant temperature gradient. In the latter case, we used the idealized profile shown in red in Fig.\ref{fig:Temperature_profile}, using the four temperature values given by the experimental data. The typical thickness of the boundary layer is set at 3 mm. The resolution of the simulations reported in the main text is $Nr\times Nz=300\times300$.

}

\showmatmethods{} 

\acknow{
We are grateful to L. Bonnet, N. Garroum, A. Leclercq, and P. Pace for their technical support and we thank S. Ismael and M. Sardin for machining the experiment. We also thank Martin Caelen, Basile Gallet and Francois Petrelis for their insightful discussions.
CG acknowledges financial support from the French program JCJC managed by Agence Nationale de la Recherche (Grant ANR-19-CE30-0025-01) and the Institut Universitaire de France.}

\showacknow{} 

\bibsplit[2]

\bibliography{TE_bib2}
\onecolumn

\begin{Huge}\begin{center} Supporting Information\end{center} \end{Huge}

\section*{Sidewall convection}
The presence of horizontal temperature gradient naturally leads to sidewall convection which appears at non-zero $\Delta T_0$. The Rayleigh number $Ra = \alpha \Delta T_0 \Delta R^3/\kappa \nu$ where $\alpha$ is the thermal expansion coefficient, $\Delta T_0$ the temperature difference between the cylinders, $\Delta R = R_o - R_i$, $\kappa$ the thermal diffusivity and $\nu$ the kinematic viscosity. For liquid Gallium, $\alpha = 5.5\cdot 10^{-5}~K^{-1}$, $\kappa = 1.3\cdot 10^{-5}~m^2.s^{-1}$, $\nu=3.18\cdot 10^{-7}~m^2.s^{-1}$. The Rayleigh number for $\Delta T_0 \sim 2-37~K$ is $Ra_{Ga}\sim 5.7\cdot 10^3 - 1.06\cdot 10^5$. For liquid Mercury, $\alpha = 1.83\cdot 10^{-4}~K^{-1}$, $\kappa = 4.9\cdot 10^{-6}~m^2.s^{-1}$, $\nu=1.49\cdot 10^{-7}~m^2.s^{-1}$. The Rayleigh number for $\Delta T_0 \sim 2-37~K$ is $Ra_{Hg}\sim 1.08 - 20.03\cdot 10^5$.

\section*{Analytical model}

We derive here a simple analytical model describing the generation of a thermoelectric current, the corresponding magnetic field, and electric potential, in a rectangular domain made of two dissimilar metals. The two electrically conducting regions, denoted by the indices $'+'$ or $'-'$, have  electrical conductivity $\sigma^{\pm}$ and Seebeck coefficient (or thermoelectric power) $S^{\pm}$. Both are supposed independent of temperature. A horizontal thermal gradient of arbitrary shape is applied across the two metals, which are separated by an electrically conducting interface located at $z=0$.

\begin{figure}[h!]
    \centering
    \includegraphics[scale=0.4]{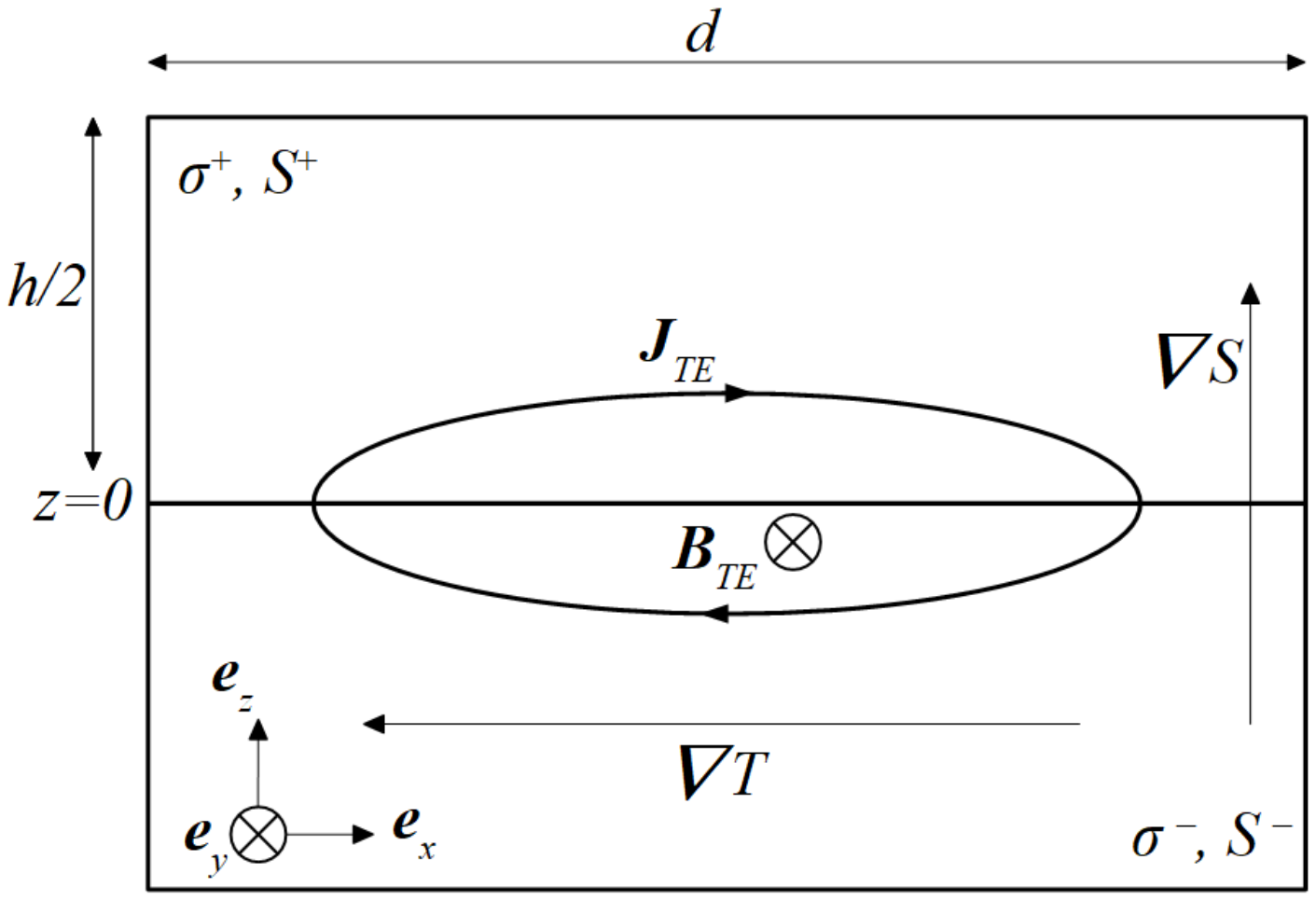}
    \caption{Two metals with Seebeck coefficients $S^{\pm}$ and electrical conductivities $\sigma^{\pm}$, superimposed in a rectangular closed domain, are in electrical contact at $z=0$, and subjected to a horizontal temperature gradient.}
    \label{fig:scheme_Bte}
\end{figure}

In the absence of a velocity field $\bm{u}$ and in the presence of a thermal gradient, Ohm's law reads: 
\begin{align}
    \frac{\bm{j}}{\sigma} = \bm{E} - S\bm{\nabla}T,
    \label{eq:Ohm_law}
\end{align}
where $\bm{j}$ is the electric current density, $\sigma$ is the electrical conductivity, $\bm{E}$ is the electric field, $S$ is the Seebeck coefficient and $T$ is the temperature field. 
%
%

\tcr{In the following we will use the magnetostatic approximation, relatively well satisfied here: in liquid metal, the magnetic field generally evolves on time scales much smaller than all the other variables such as the temperature or the velocity field. This is summed up by the dimensionless number $\zeta = \mu_0\sigma \kappa$, with $\mu_0$ the vacuum magnetic permeability. $\zeta$ is the ratio of the temperature evolution time scale due to thermal diffusion to the magnetic evolution time scale (also due to diffusion). The presence of convection implies that the temperature can evolve on time scale faster than $\Delta R^2/\kappa$ like the eddy turnover time, $\Delta R/U_{ff}$ and $U_{ff}$ being a typical velocity scale due to convection such as the free-fall velocity $U_{ff}\sim \sqrt{\alpha\Delta T_0 gh}$. In that case, $Rm=\mu_0\sigma U_{ff}\Delta R$ must also be small to fulfill the quasi-static approximation. In the present experiment, both $\zeta\ll 1$ and $Rm\ll 1$, ensure that the evolution of the magnetic field produced by thermoelectricity follows adiabatically the evolution of temperature.}

In the magnetostatic approximation and for steady state, the Maxwell-Faraday equation reads $\bm{\nabla}\times\bm{E} = 0$. For each layer, the electric field can then be decomposed as follows, $\bm{E} = -\bm{\nabla}V^\pm$ where $V^\pm$ is the electric potential in each subdomain.

\tcr{Taking the curl of the Ohm's law (\ref{eq:Ohm_law}) in each subdomain: 
\begin{align}
\nabla\times \left(\frac{\bm{j}^\pm}{\sigma^\pm}\right) = -\nabla\times \left(S\bm{\nabla}T\right)=\bm{\nabla} S\times \bm{\nabla} T
    \label{eq:Ohm_law2}
\end{align}
Because $S(T)$ is a function of temperature only, $\bm{\nabla} S\times \bm{\nabla} T=0$. With the assumption that the electrical conductivity is  constant in each domain, we get :
\begin{align}
\bm{j^\pm}=-\sigma^\pm\bm{\nabla}\phi^\pm
    \label{eq:phi}
\end{align}}

The charge conservation, in the magnetostatic approximation, implies $\bm{\nabla}\cdot\bm{j}^\pm = 0$. Therefore, in each domain, $\phi^\pm$ fulfills a Laplace equation $\nabla^2\phi^\pm = 0$. The boundary conditions for the current are prescribed by charge conservation:
\begin{align}
	j_x^\pm(x=0,z) = j_x^\pm(x=d,z) = 0, \\ j_z^+(x,z=h/2) = j_z^-(x,z=-h/2) = 0, \\ j_z^+(x,z=0^+) = j_z^-(x,z=0^-)
\end{align}
These boundary conditions can be translated for  $\phi^\pm$ as:
\begin{align}
	\partial_x\phi^\pm(x=0,z) = \partial_x\phi^\pm(x=d,z) = 0, \\ \partial_z\phi^+(x,z=h/2) = \partial_z\phi^-(x,z=-h/2) = 0, \\ \sigma^+\partial_z\phi^+(x,z=0^+) = \sigma^-\partial_z\phi^-(x,z=0^-)
\end{align}
The quantity $\phi^\pm$ can then be obtained as a decomposition over the eigenfunctions of the Laplacian. It is clear that $\sin(n\pi x/d)$, with $n\in \mathbb{N}$, fulfill the boundary conditions for $\partial_x\phi^\pm$, thus
\begin{align}
	\phi^\pm = \sum_n \cos\left(\frac{n\pi x}{d}\right)g_n^\pm(z).
\end{align}
As $\phi^\pm$ respects a Laplace equation, it is easy to check that $g_n^\pm(z)= a_n^\pm \cosh(\kappa_n z) + b_n^\pm \sinh(\kappa_n z)$ with $\kappa_n = n\pi /d$ for simplicity. The boundary conditions at $z = \pm h/2$ then implies: 
\begin{align}
	\frac{dg_n^{\pm}}{dz} (z=\pm h/2) = \kappa_n a_n^\pm \sinh(\pm\kappa_n h/2) + \kappa_n b_n^\pm\cosh(\pm\kappa_n h/2) = 0,
\end{align}
which is a constraint on the coefficients since $b_n^\pm = \mp \tanh(\kappa_n h/2) a_n^\pm$. Injected in  $\phi^\pm$, it gives:
\begin{align}
	\phi^\pm = \sum_n a_n^\pm \cos(\kappa_n x)(\cosh(\kappa_n z) \mp \tanh(\kappa_n h/2)\sinh(\kappa_n z)).
\end{align}
Finally, the boundary condition at $z=0$ for $\phi^\pm$ links the coefficients $a_n^+$ and $a_n^-$. Indeed, it is easy to check that $a_n^- = -\sigma^+ a_n^+/\sigma^-$. 
The continuity of the electric potential at the interface between the two conductors gives:
\begin{align}
	V^+(x,z=0^+) - V^-(x,z=0^-) = 0,
\end{align}

\tcr{Using the Ohm's law $\bm{\nabla} V^\pm=\bm{\nabla} (\phi^\pm-S^\pm T)$ where $S$ is considered constant in each phase, the previous expression can be recast in terms of $\phi^\pm$:}
%
\begin{align}
	\phi^+(x,z=0^+) - \phi^-(x,z=0^-) = \Delta S T(x,0),
\end{align}
with $\Delta S = S^+ - S^-$. Injecting the expression of $\phi^+$ and $\phi^-$ gives: 
\begin{align}
	\sum_n a_n^+ \frac{\sigma^+ + \sigma^-}{\sigma^-}\cos(\kappa_n x) = \Delta S T(x,z=0),
\end{align}
multiplying this expression by $\cos(\kappa_m x)$ and integrating over the interval $[0,d]$ enables to obtain the expression of $a_n^+$ (where the orthogonality relation for trigonometric function has been used): 
\begin{align}
	a_n^+ = \frac{K_n \sigma^- \Delta S}{ d (\sigma^+ + \sigma^-)}\int_0^d T(x,0)\cos(\kappa_n x)dx.
\end{align}
with $K_n = 1 $ if $n=0$ and $K_n = 2$ otherwise. Finally, this gives the potential:
\begin{align}
	\phi^\pm = \pm \sum_n \frac{K_n \sigma^\mp\Delta S}{d (\sigma^+ + \sigma^-)}\cos(\kappa_n x)(\cosh(\kappa_n z) \mp \tanh(\kappa_n h/2)\sinh(\kappa_n z))\int_0^d T(x,0)\cos(\kappa_n x)dx.
\end{align}
The potential $\phi$ which prescribes the thermoelectric current distribution is therefore completely determined by the temperature profile at the interface. The computation of $\bm{j}^\pm$ and $\bm{B}$ which is given by Maxwell-Ampère law's $\bm{\nabla}\times\bm{B} = \mu_0\bm{j}$, is straightforward: 
\begin{align}
	j_x^\pm = \pm \sum_n \frac{K_n \tilde{\sigma}\Delta S \kappa_n}{ d }\sin(\kappa_n x)(\cosh(\kappa_n z) \mp \tanh(\kappa_n h/2)\sinh(\kappa_n z)) I_n(T),\\
	j_z^\pm = \mp \sum_n \frac{K_n \tilde{\sigma}\Delta S \kappa_n}{ d }\cos(\kappa_n x)(\sinh(\kappa_n z) \mp \tanh(\kappa_n h/2)\cosh(\kappa_n z)) I_n(T),
\end{align}
with $ \tilde{\sigma} = \sigma^+\sigma^-/(\sigma^+ + \sigma^-)$ and $I_n(T) = \int_0^d T(x,0)\cos(\kappa_n x)dx$. The important point of this result is the fact that any variation of the temperature along $z$ will be supported by $V$ keeping $\phi$, $\bm{j}$, and $\bm{B}$ unchanged. The component of the magnetic field produced by the thermoelectric effect is orthogonal to the plane $(x,z)$, $B_y$ simply denoted $B$ and is: 
\begin{align}
	B^\pm = \mp \sum_n \frac{K_n \mu_0 \tilde{\sigma}\Delta S }{ d}\sin(\kappa_n x)(\sinh(\kappa_n z) \mp \tanh(\kappa_n h/2)\cosh(\kappa_n z)) I_n(T),
\end{align}

\begin{figure}[h!]
    \centering
    \includegraphics[scale=0.8]{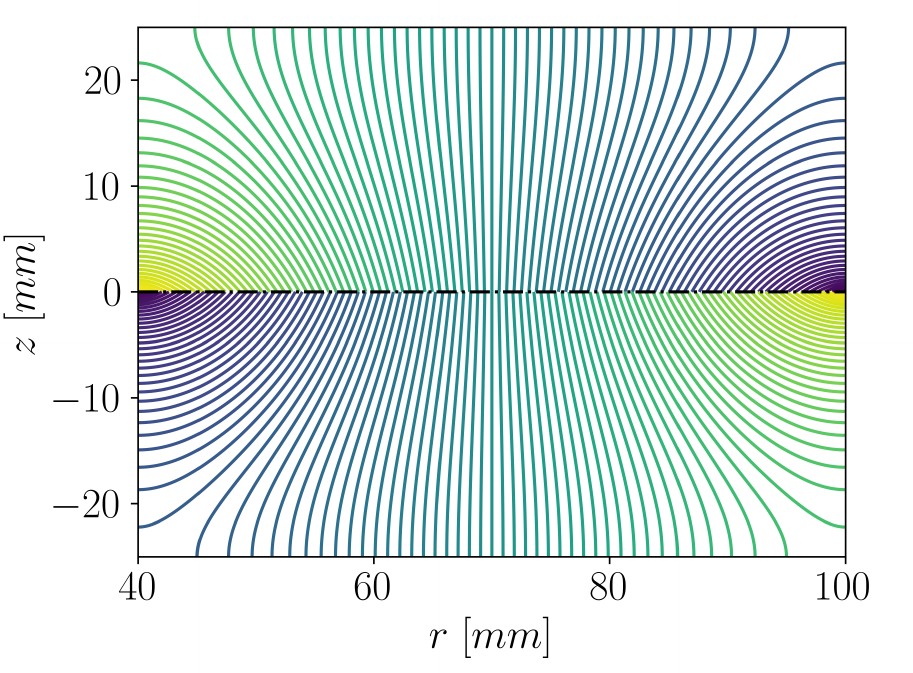}
    \caption{Line of potential $\phi$ in the cartesian domain $[0,d]\times [-h/2,h/2]$. The dashed-dotted line corresponds to the position of the interface. The temperature profile at the interface displays a linear gradient, corresponding to the case where at least one of the metals is solid.}
    \label{fig:solide_phi_colormap}
\end{figure}
\begin{figure}[h!]
    \centering
    \includegraphics[scale=0.8]{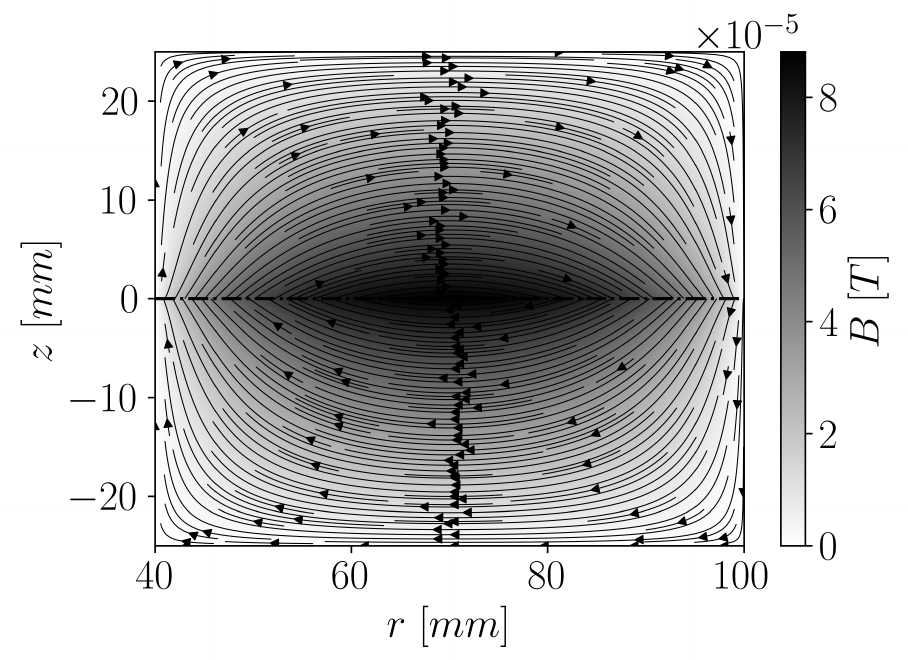}
    \caption{Colormap of the magnetic field $B$ in the cartesian domain $[0,d]\times [-h/2,h/2]$. The dashed-dotted line corresponds to the position of the interface. The black lines are the electric current. The temperature profile at the interface displays a linear gradient, corresponding to the case where at least one of the metals is solid.}
    \label{fig:solide_B_colormap}
\end{figure}

We now implement this expression using the geometry and properties of the metals used in the experiment, namely mercury and gallium, $h=25$ mm, $d=60$ mm. If the two metals were in a solid state, the temperature profile would be linear with a constant thermal gradient $-\Delta T_0/d$, where $\Delta T_0$ is the thermal gradient applied at the horizontal wall boundaries. Fig. \ref{fig:solide_phi_colormap} shows the computed isoline of potential $\phi^\pm$ while Fig. \ref{fig:solide_B_colormap} shows a colormap of $B$ for $n_{max} = 400$, using the value $\Delta T_0 = 37$K obtained in the experiment at maximum heating power. The black lines correspond to the streamlines of the thermoelectric current. The resolution used to plot the solution is $dx=5\cdot 10^{-4} d $ and $dz=5\cdot 10^{-4} h$.\\

\begin{figure}[h!]
    \centering
    \includegraphics[scale=0.8]{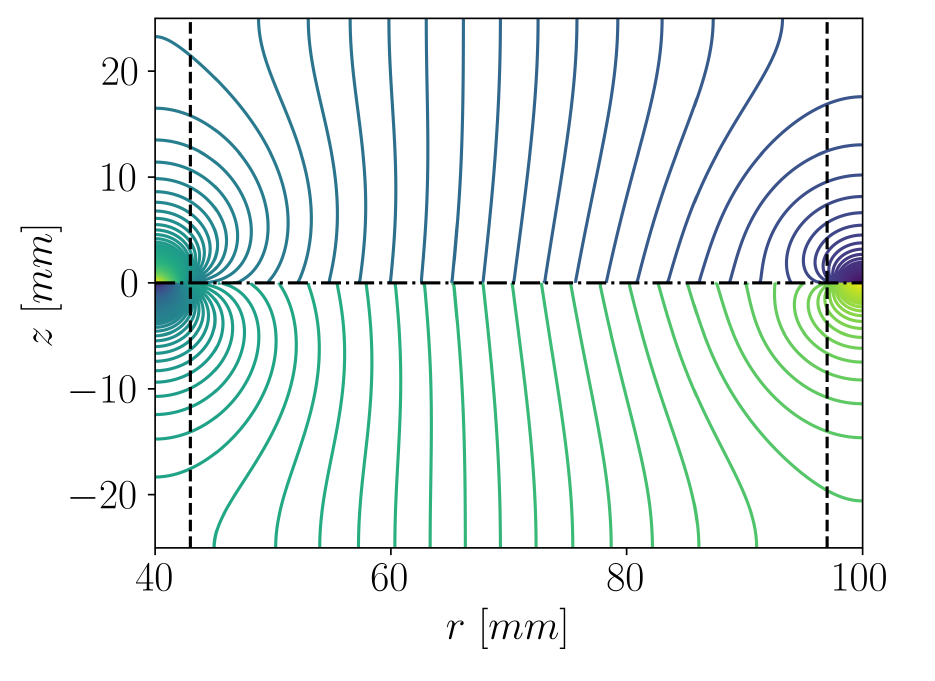}
    \caption{Line of potential $\phi$ in the cartesian domain $[0,d]\times [-h/2,h/2]$. The dashed-dotted line corresponds to the position of the interface. The temperature profile at the interface is a piecewise linear gradient, and the vertical dashed lines indicate the positions of the thermal boundary layers..}
    \label{fig:2BL_phi_colormap}
\end{figure}

\begin{figure}[h!]
    \centering
    \includegraphics[scale=0.8]{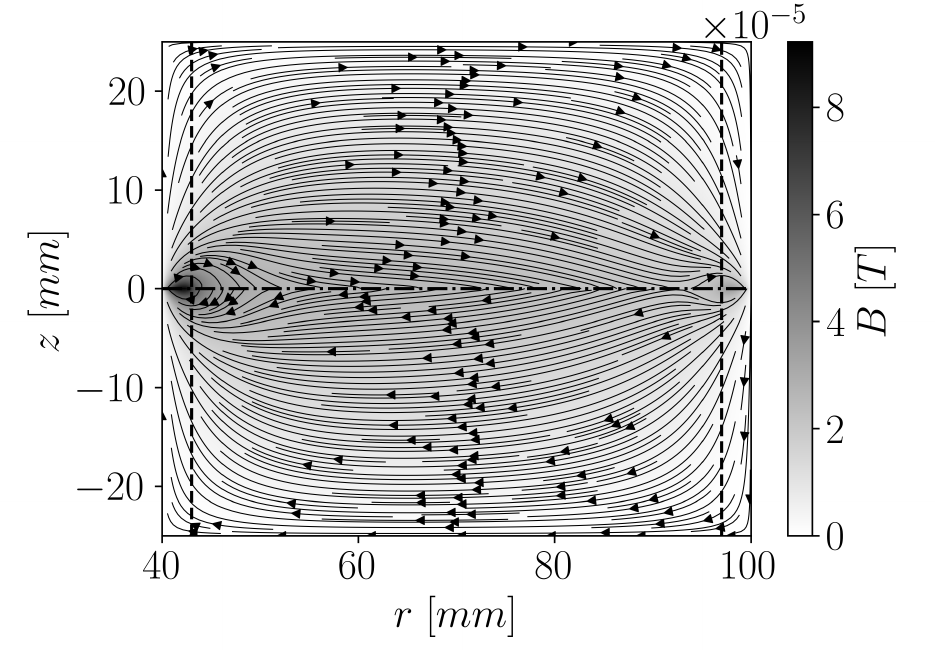}
    \caption{Colormap of the magnetic field $B$ in the cartesian domain $[0,d]\times [-h/2,h/2]$. The dashed-dotted line corresponds to the position of the interface. The black lines are the electric current. The temperature profile at the interface is a piecewise linear gradient, and the vertical dashed lines indicate the positions of the thermal boundary layers.}
    \label{fig:2BL_B_colormap}
\end{figure}

In the more realistic case of an interface separating two liquid metals, as in the experiment, the temperature profile can be approximated as piecewise linear at the interface. Here again, we use the temperatures obtained in the experiment (the red profile shown in Fig.2 of the main text). The resulting solution is shown in Fig \ref{fig:2BL_phi_colormap} and Fig \ref{fig:2BL_B_colormap}. The results are in excellent agreement with those obtained from the direct numerical simulations reported in the main manuscript, and confirm the existence of intense current loops near the boundaries and a saddle point at the interface.

\begin{figure}[h!]
    \centering
    \includegraphics[scale=0.5]{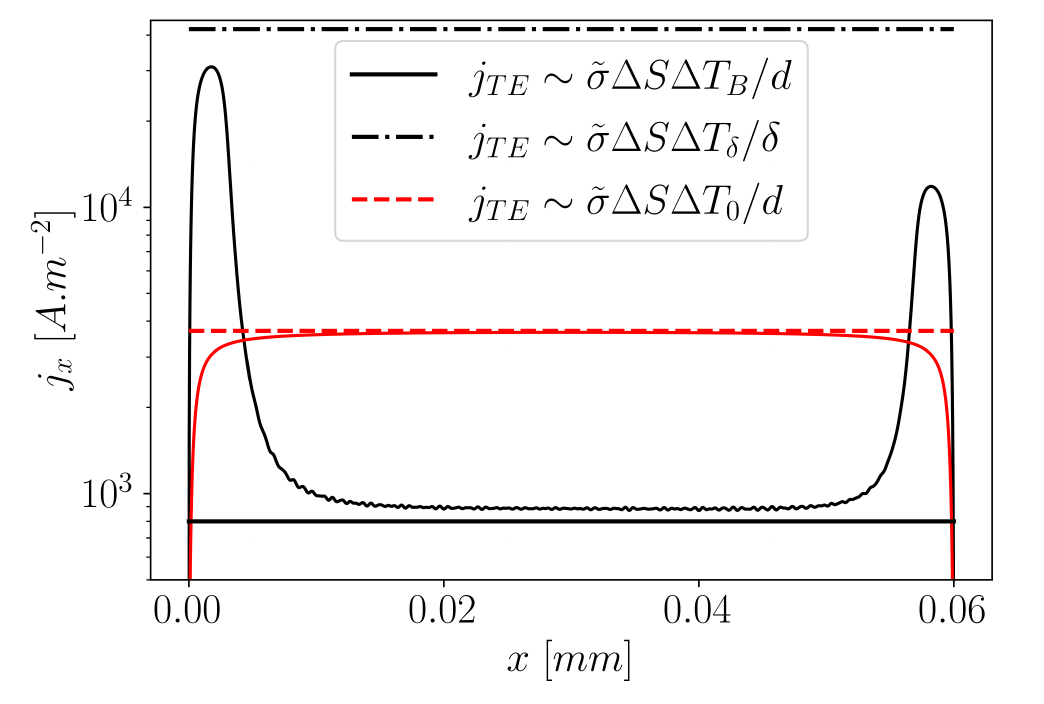}
    \caption{Comparison between the horizontal component of the thermoelectric current density for a solid (red line) and a liquid interface (black line) both taken at $z=+0.5mm$.}
    \label{fig:comparison}
\end{figure}

 Fig. \ref{fig:comparison} shows the horizontal component of the thermoelectric current at $z=+0.5mm$ for the two cases studied. Far enough from the vertical walls, a good estimate of $j_x$ in the solid case is $\Tilde{\sigma}\Delta S\Delta T_0/d$ while for the liquid case, $\Tilde{\sigma}\Delta S\Delta T_B/d$ provides the correct estimate, in agreement with numerical predictions.

This agreement between theoretical predictions and numerical results confirms that the geometry of thermoelectric currents and magnetic field strength are controlled by the temperature profile at the interface, $\Tilde{\sigma}$ and $\Delta S$. This also confirms that the liquid nature of the interface, which produces a complex non-linear temperature profile, can generate a non-trivial distribution of thermoelectric currents, particularly near the thermal boundaries.\\


\end{document}